\documentclass[twocolumn]{jpsj2}
\def\gsim{\mathop {\vtop {\ialign {##\crcr 
$\hfil \displaystyle {>}\hfil $\crcr \noalign {\kern1pt \nointerlineskip
 } 
$\,\sim$ \crcr \noalign {\kern1pt}}}}\limits}
\def\lsim{\mathop {\vtop {\ialign {##\crcr 
$\hfil \displaystyle {<}\hfil $\crcr \noalign {\kern1pt \nointerlineskip
 } 
$\,\,\sim$ \crcr \noalign {\kern1pt}}}}\limits}

\def\runauthor{
}
\title{Effect of Spin-Orbit Interaction on (4$d)^{3}$- and (5$d)^{3}$-Based Transition-Metal Oxides}          
\author{ Hiroyasu {\sc Matsuura}$^{1}$ and Kazumasa {\sc Miyake}$^{2}$ 
}
\inst
{
$^{1}$Department of Physics, University of Tokyo, 7-3-1 Hongo, Bunkyo, Tokyo 113-0033, Japan\\
$^{2}$Toyota Physical and Chemical Research Institute,
41-1 Yokomichi, Nagakute, Aichi 480-1192, Japan\\
}

\recdate
{
\today
}

\abst
{The effects of a spin-orbit interaction on transition-metal ions of (4$d$)$^3$- and (5$d$)$^3$-based oxides in which three electrons occupy t$_{2{\rm g}}$ orbitals are studied. 
The amplitude of the magnetic moment of $d$ electrons on the 5$d$ and 4$d$ orbitals is estimated by numerical diagonalization. 
It is found that the magnetic moment is reduced by the spin-orbit interaction.
It is suggested that (4$d)^3$- and (5$d)^3$-based oxides are located in the middle of the L-S and J-J coupling schemes. 
}

\kword{(4$d)^3$ electron, (5$d)^3$ electron, spin-orbit interaction, L-S coupling, J-J coupling 
}
\begin{document}
\sloppy

\maketitle

\begin{figure}[h]
\begin{center}
\rotatebox{0}{\includegraphics[angle=-90,width=1\linewidth]{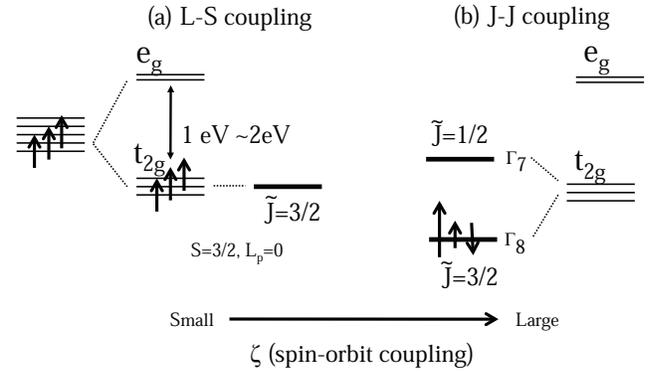}}
\caption{Schematic pictures of (a) L-S coupling and (b) J-J coupling in $d^{3}$ state.}
\label{Fig1}
\end{center}
\end{figure}

Transition-metal oxides, which have three 5$d$ or 4$d$ electrons, have been focused on recently.
For example, NaOsO$_3$ (5$d^3$) exhibits a novel metal-insulator transition called the Slater transition~\cite{Shi}, and SrTcO$_3$ (4$d^3$) has a very high antiferromagnetic transition temperature of $T_N \sim 1000$ K~\cite{Rodriguez}.
 Since the Os ion (or Tc ion) is surrounded by six oxygens, the five 5$d$ (or 4$d$) orbitals are split into the t$_{2{\rm g}}$ and e$_{\rm g}$ orbitals by the crystalline electric field (CEF), as shown in Fig. \ref{Fig1}(a).
Thus, three electrons occupy t$_{2{\rm g}}$ orbitals.

When we study a spin state of $d^3$ electrons, we usually adopt the L-S coupling scheme particularly in the 3$d$ case: we construct the $S=3/2$ and $L_p=0$ states using Hund's rule as shown in Fig. \ref{Fig1}(a), where $S$ and $L_p$ are the total spin and orbital angular momenta in the $t_{2\rm g}$ orbital.
It is thought that the spin-orbit interaction (SOI) does not play an important role in determining the electronic state of the $d^3$ system because $L_p=0$.
Thus, it is expected that the amplitude of the magnetic moment will be about 3 $\mu_{\rm B}$.
   
Recently, it has been reported that the amplitude of the magnetic moment of the Os ion in NaOsO$_3$ is about 1.0 $\mu_{\rm B}$, as determined by the neutron scattering experiment~\cite{Calder}.
The electronic state of NaOsO$_3$ has been discussed from the results of the first principles calculations, and the amplitude is consistent with the experimental value~\cite{Du,Jung}.
It was also reported that the amplitude of the magnetic moment of the Tc ion in SrTcO$_3$ is about 1.9 $\mu_{\rm B}$~\cite{Rodriguez}.
This electronic state has also been studied using first principles calculations and the effective model, and the amplitude has been found to be consistent with the experimental value~\cite{Mravlje,Middey}.
This magnetic moment has been discussed on the basis of the idea that the reduction in the magnetic moment is due to the hybridization between the $d$ and $p$ orbitals around a transition metal.
However, the origin of the reduction in the magnetic moment has not yet been fully understood.
In particular, the role of SOI has not been well studied in these $d^3$ systems.  

In this Letter, we show that the magnetic moments of (4$d)^{3}$ and (5$d)^{3}$ ions are reduced by SOI. Then, we find that these $d^3$ ions are located in the middle of the L-S and J-J coupling schemes. 

To understand the electronic state of $d^{3}$ ions, we use the following Hamiltonian:
\begin{eqnarray}
H_{{\rm tot}} = H_{d}^{{\rm int}} + H^{{\rm SO}}_{{d}}, \label{Ham}
\end{eqnarray}
where $H_{{d}}^{{\rm int}}$ and $H^{{\rm SO}}_{{d}}$ are the Hamiltonians for Coulomb interactions and SOI on $d$ orbitals.
Their explicit forms are given as  
\begin{eqnarray}
H_{d}^{{\rm int}}  &=&  U_d \sum_{i=1,2,3}n_{i\uparrow}n_{i\downarrow} \nonumber \\
             & & +\frac{U_{d}^{\prime} - J_{d}}{2}\sum_{ \substack{ i,i^{\prime}=1,2,3 \\ (i \neq i^{\prime}) } }\sum_{\sigma} n_{i\sigma}n_{i^{\prime}\sigma} \nonumber \\  
      & & +\frac{U_{d}^\prime}{2} \sum_{\sigma\neq \sigma^\prime} \sum_{\substack{ i,i^{\prime}=1,2,3 \\ (i \neq i^{\prime}) }} n_{i\sigma}n_{i^{\prime}\sigma^{\prime}}  \nonumber\\ 
 &&+ \frac{J_{d}}{2}\sum_{\substack{ i,i^{\prime}=1,2,3 \\ (i \neq i^{\prime}) }}( d_{i^{\prime}\uparrow}^{\dag}d_{i\uparrow }d_{i\downarrow }^{\dag}d_{i^{\prime}\downarrow } + {\rm h.c.}),\\
H^{{\rm SO}}_{{\rm d}} &=& \frac{i\zeta}{2} \sum_{lmn} \epsilon_{lmn}\sum_{\sigma,\sigma^{\prime} } \sigma_{\sigma\sigma^{\prime}}^{n} d_{i,l\sigma}^{\dagger}d_{i,m\sigma^{\prime}},
\end{eqnarray} 
where $U_d$, $U_d^{\prime}$, $J_{d}$, and $\zeta$ are the intra- and inter-Coulomb interactions, ferromagnetic exchange interaction (Hund's rule coupling), and the magnitude of SOI between {\bf l}$_i$ and {\bf s}$_i$ where {\bf l}$_i$ and {\bf s}$_i$ are the orbital and spin angular momenta of the $i$-th electron, respectively.
These Coulomb interactions have the relation $U_{d}= U_{d}^{\prime} + 2J_d$. 
$d_{i\sigma}$ is the annihilation operator of the $i$-th orbital ($i=1,2,3$) with a spin $\sigma$, $n_{d} \equiv d_{i\sigma}^{\dagger}d_{i\sigma}$, and $\epsilon_{lmn}$ is the Levi-Civita symbol.   

\begin{figure}[h]
\begin{center}
\rotatebox{0}{\includegraphics[angle=0,width=1\linewidth]{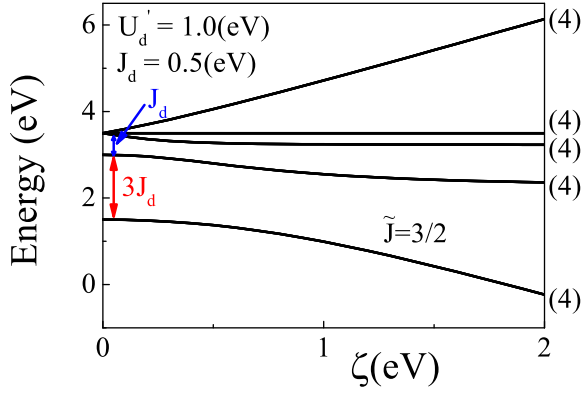}}
\caption{(Color online) Spin-orbit interaction ($\zeta$) dependences of eigenvalues for $U_{d}^{\prime}=1.0$ (eV) and $J_{d}= 0.5$ (eV). The number in parentheses indicates the degree of degeneracy. }
\label{Fig2}
\end{center}
\end{figure}

By diagonalizing $H_{\rm tot}$, the eigenvalues and eigenstates are determined.
Figure \ref{Fig2} shows the $\zeta$ dependences of eigenvalues for $U_{d}^{\prime}=1.0$ (eV) and $J_{d}=0.5$ (eV).
The lowest-energy state is quadruply degenerate and characterized by the total angular momentum of $\tilde{J}=3/2$, which is a good quantum number of the Hamiltonian eq.(\ref{Ham}), where $\tilde{{\bf J}} \equiv {\bf L}_p + {\bf S}$ .
The energy of this state decreases as $\zeta$ increases.
It is also found that this energy level does not cross each other.
Note that all of the diagonal matrix elements of eq. (\ref{Ham}) are $3U^{\prime}_d -J_d$.
Since $U_{d}^{\prime}$ appears only in the diagonal matrix elements, the energy difference between eigenvalues is independent of $U^{\prime}_d$.

\begin{figure}[h]
\begin{center}
\rotatebox{0}{\includegraphics[angle=0,width=1\linewidth]{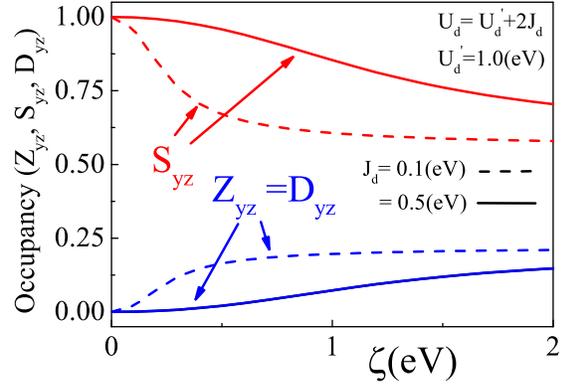}}
\caption{(Color online) Zero, single, and double occupancy on yz orbital for $J_{d}=0.1$(eV) (dashed line) and $J_{d}=0.5$(eV) (solid line). 
Red and blue lines indicate single and zero (double) occupancy. The line of zero occupancy is identical with that of double occupancy.}
\label{Fig3}
\end{center}
\end{figure}

Figure \ref{Fig3} shows the $\zeta$ dependences of the zero, single, and double occupancies of the $yz$ orbital for $J_{d}=0.1$ (eV) (dashed lines) and $0.5$ (eV)(solid lines). 
Here, the occupancy of the $yz$ orbital in the ground state is estimated as
\begin{eqnarray}
O_{yz} = \frac{1}{4}\sum_{j=1}^{4}\langle j|\hat{O}_{yz}|j \rangle,
\end{eqnarray}  
where $j$ indicates quadruply degenerate ground states, and the operators of the zero, single, and double occupancies are respectively defined by
\begin{eqnarray}
\hat{Z}_{yz} &\equiv&  1 -n_{1\uparrow} - n_{1\downarrow} + n_{1\uparrow}n_{1\downarrow} , \\
\hat{S}_{yz} &\equiv&  n_{1\uparrow} + n_{1\downarrow} -2n_{1\uparrow}n_{1\downarrow} , \\ 
\hat{D}_{yz} &\equiv&  n_{1\uparrow}n_{1\downarrow}.
\end{eqnarray}  
Since the split due to the CEF is zero, ${Z}_{yz}={Z}_{zx}={Z}_{xy}$, ${S}_{yz}={S}_{zx}={S}_{xy}$, and ${D}_{yz}={D}_{zx}={D}_{xy}$.  
The single occupancy is indicated by red lines. 
The zero and double occupancies are indicated by blue lines, and the line of the zero occupancy is identical to the line of the double occupancy, because these states of the zero and double occupancies appear simultaneously.
It is found that the zero and double occupancies increase as $\zeta$ increases.
It is suggested that the decrease in $S_{yz}$ denotes the reduction in the spin moment, while the increase in $Z_{yz}$ and $D_{yz}$ indicates the enhancement in the orbital moment.

\begin{figure}[h]
\begin{center}
\rotatebox{0}{\includegraphics[angle=0,width=1\linewidth]{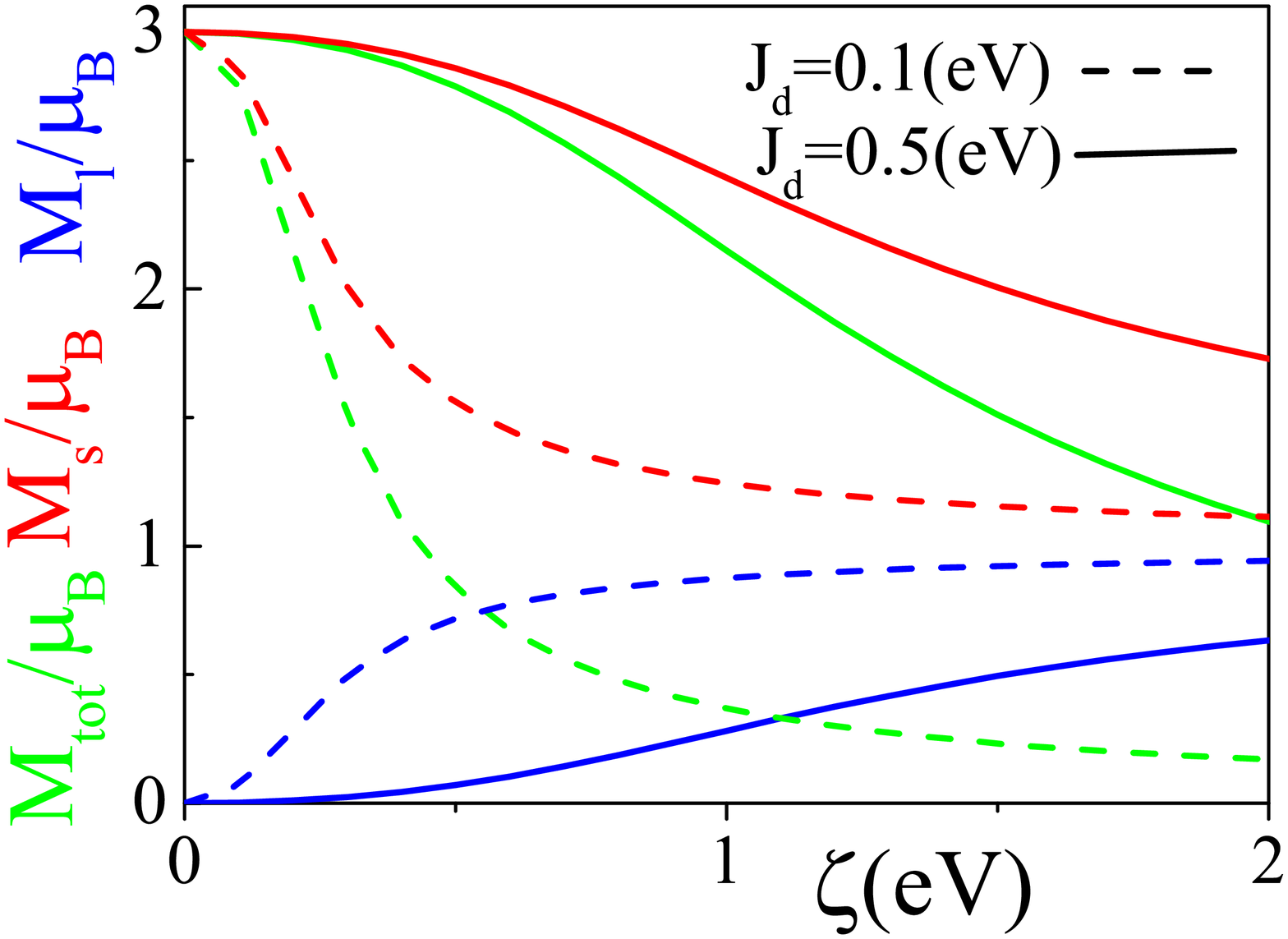}}
\caption{(Color Online) The amplitudes of the total, spin, and orbital magnetic moments ($M_{\rm tot}/\mu_{\rm B}$, $M_s/\mu_{\rm B}$, $M_l/\mu_{\rm B}$) are for $J_{d}=0.1$(eV) (dashed line) and $J_{d}=0.5$(eV) (solid line), which are expressed by the green, red, and blue lines, respectively.
}
\label{Fig4}
\end{center}
\end{figure}

Figure \ref{Fig4} shows the $\zeta$ dependence of the amplitude of the magnetic moment ($M_{\rm tot}/\mu_{\rm B}$, $M_s/\mu_{\rm B}$, $M_l/\mu_{\rm B}$) for $J_{d}=0.5$ (eV) (solid lines) and $J_{d}=0.1$ (eV) (dashed lines), where $M_{\rm tot}/\mu_{\rm B}$, $M_s/\mu_{\rm B}$, and $M_l/\mu_{\rm B}$ are the amplitudes of the total, spin, and orbital magnetic moments, respectively.
$M_{\rm tot}/\mu_{\rm B}$, $M_s/\mu_{\rm B}$, and $M_l/\mu_{\rm B}$ are defined by $M_{\rm tot}/\mu_{\rm B}= |\sum_{i}(l_z^i +2s_z^i)|$, $M_s/\mu_{\rm B}= |\sum_{i}2 s_z^i|$, and $M_l/\mu_{\rm B}= |\sum_{i} l_z^i|$, where $s_z^i$ and $l_z^i$ are the $z$ components of the spin and orbital angular momenta of the $i$-th electron, respectively.
These are expressed by the green, red, and blue lines, respectively.
As $\zeta$ increases, $M_s/\mu_{\rm B}$ decreases, while $M_l/\mu_{\rm B}$ increases.
This is because the zero and double occupancies increase, as shown in Fig. \ref{Fig3}.
Then, $M_{\rm tot}/\mu_{\rm B}$ decreases in total.
The limiting values of $M_l/\mu_{\rm B}$ and $M_s/\mu_{\rm B}$ for $\zeta \rightarrow \infty$ with $J_d=0.1$ (eV) (shown in Fig. \ref{Fig4}) suggest that $L_p$ and $S$ in the limit of $\zeta \rightarrow \infty$ are $L_p=1$ and $S=1/2$, because $M_l/\mu_{\rm B} = L_p^z = 1.0$ and $M_s/\mu_{\rm B} = 2S^z = 1.0$ where $L_p^z$ and $S^z$ are the z components of ${\bf L}_p$ and ${\bf S}$, respectively.

The reason why $M_{\rm tot}/\mu_{\rm B}$ decreases is explained by the J-J coupling scheme, as shown in Fig. \ref{Fig1}(b).
Since the angular momentum of t$_{2{\rm g}}$ orbitals, ${\bf l}_{{\rm t}_{2{\rm g}}}$, corresponds to that of pseudo $p$ orbitals ${\bf l}_{p}$ as ${\bf l}_{{\rm t}_{2{\rm g}}} = - {\bf l}_{p}$, 
the $\Gamma_{8}$ quartet and $\Gamma_7$ doublet are constructed by SOI between ${\bf s}=1/2$ and ${\bf l}_p=1$.
Three electrons on the t$_{2{\rm g}}$ orbital occupy $\Gamma_8$ states of $\tilde{j}_z=3/2$, $1/2$, and $-1/2$ as shown in Fig. \ref{Fig1}(b), where $\tilde{{\bf j}} \equiv {\bf l}_{p} +{\bf s}$.
Since the $d^3$ state corresponds to the case of one hole on $\Gamma_8$ orbitals, its situation is the same as that in the case of one electron on $\Gamma_8$ orbitals.    
It is well known that the Land\'{e} g factor of $\tilde{j}=3/2$ on the $t_{2{\rm g}}$ orbital, which may be called a ``modified'' Land\'{e} g factor, is $\tilde{g}_{\tilde{j}=3/2}=0$ for $\zeta \rightarrow \infty$, if one electron occupies $\Gamma_8$ and the effect of $\Gamma_7$ is neglected~\cite{Kamimura}.
Thus, it is found that the g factor in the case of one hole is also $\tilde{g}_{\tilde{J}=3/2} =0$.

\begin{figure}[h]
\begin{center}
\rotatebox{0}{\includegraphics[angle=0,width=0.5\linewidth]{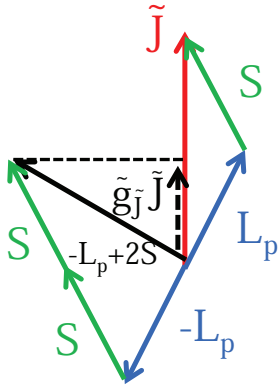}}
\caption{(Color Online) Schematic picture of ${\bf L}_{p}$, ${\bf S}$, $\tilde{{\bf J}}$, and $\tilde{g}_{\tilde{J}}\tilde{{\bf J}}$, where $\tilde{g}_{\tilde{J}}$ is a ``modified'' Land\'{e} g factor and $\tilde{g}_{\tilde{J}}\tilde{{\bf J}}$ is the component of $-{\bf L}_{p}+2{\bf S}$ parallel to $\tilde{{\bf J}}$.
}
\label{Fig5}
\end{center}
\end{figure}

Generally, the Land\'{e} g factor can be estimated independently of the amplitude of $\zeta$.
Figure \ref{Fig5} shows a schematic picture of $\tilde{{\bf J}}$, ${\bf L}_p$, ${\bf S}$, $-{\bf L}_p +2{\bf S}$, and $\tilde{g}_{\tilde{J}} \tilde{{\bf J}}$, which is the component of $-{\bf L}_p +2{\bf S}$ parallel to $\tilde{{\bf J}}$.
Then, the component of ${\bf M}$ parallel to $\tilde{{\bf J}}$, where ${\bf M} \equiv \mu_{\rm B}(-{\bf L}_p +2{\bf S})$, is determined by the relations 
\begin{eqnarray}
{\bf M} \cdot \tilde{{\bf J}} = \tilde{g}_{\tilde{J}} \mu_{\rm B} \tilde{{\bf J}}^2, 
\end{eqnarray}
and
\begin{eqnarray}
{\bf M} \cdot \tilde{{\bf J}} = \bigr[\frac{3}{2}({\bf S}^2 -{\bf L}_p^2 ) +\frac{1}{2}\tilde{{\bf J}}^2 \bigr]\mu_{\rm B}.
\end{eqnarray}
Therefore, the ``modified'' Land\'{e} g factor is given by
\begin{eqnarray}
\tilde{g}_{\tilde{J}} = \frac{3}{2}\frac{S(S+1) -L_p(L_p+1)}{\tilde{J}(\tilde{J}+1)} +\frac{1}{2}.
\end{eqnarray}
It is found that $\tilde{g}_{\tilde{J}}=2$ and $M_{\rm tot}/\mu_{\rm B} =3.0$ for $L_p =0$, $S=3/2$, and $\tilde{J}=3/2$ corresponding to $\zeta =0$.
On the other hand, $\tilde{g}_{\tilde{J}} = 0$ and $M_{\rm tot}/\mu_{\rm B} = 0.0$ for $L_p = 1$, $S = 1/2$, and $\tilde{J} = 3/2$ corresponding to $\zeta \rightarrow \infty$, in accordance with the result shown in Fig. \ref{Fig4}.
As a result, $M_{\rm tot}/\mu_{\rm B}$ decreases toward zero as $\zeta$ increases, as shown in Fig. \ref{Fig4}.

Note that Fig. \ref{Fig4} is suitable in the case where the crystalline electric field splitting $\Delta$ between t$_{2{\rm g}}$ and e$_{\rm g}$ is much larger than SOI $\zeta$.
In the opposite limit $\Delta \ll \zeta$, the usual form of the Land\'{e} g factor is recovered, giving $g_{j}=4/5$ for $l=2$, $s=1/2$, and $j=3/2$, which is the ground state of SOI in the electron configuration of an electron on 5$d$ or 4$d$ quintuply degenerate orbitals. 
Here, ${\bf l}$ is defined as the orbital angular momentum of five d orbitals, and ${\bf j} \equiv {\bf l} +{\bf s}$.
Then, the magnetic moment of 5$d^3$ or 4$d^3$ ions becomes $M_{\rm tot}/\mu_{\rm B} =g_{j}(3/2 + 1/2 - 1/2) = 6/5$ by the J-J coupling scheme.

\begin{figure}[h]
\begin{center}
\rotatebox{0}{\includegraphics[angle=0,width=1\linewidth]{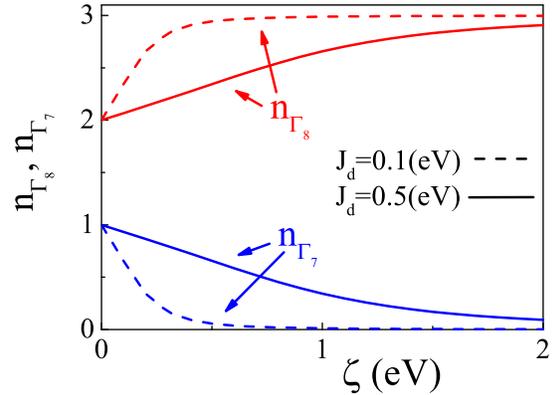}}
\caption{ (Color Online) $\zeta$ dependences of the number of electrons on four $\Gamma_8$ orbitals ($n_{\Gamma_8}$) (red line) and two $\Gamma_7$ orbitals ($n_{\Gamma_7}$) (blue line) for $J_d=0.1$ (eV) (dashed line) and $J_d=0.5$ (eV) (solid line).
}
\label{Fig6}
\end{center}
\end{figure}

The $\zeta$ dependence of the number of electrons on four $\Gamma_8$ orbitals ($n_{\Gamma_8}$) and two $\Gamma_7$ orbitals ($n_{\Gamma_7}$) in the ground state is shown in Fig. \ref{Fig6}, where the wave functions of $\Gamma_8$ and $\Gamma_7$ orbitals are defined as in ref. \citen{Kamimura}.
The red and blue lines indicate the numbers of electrons on the $\Gamma_8$ and $\Gamma_7$ orbitals, respectively.
For $\zeta=0$, the number of electrons on $\Gamma_8$ and $\Gamma_7$ orbitals ($n_{\Gamma_7}$) are  $n_{\Gamma_8}=2$ and $n_{\Gamma_7}=1$.
As $\zeta$ increases, $n_{\Gamma_8}$ increases and $n_{\Gamma_7}$ decreases, and the numbers of electrons on these orbitals approximate $n_{\Gamma_8} \simeq 3$ and $n_{\Gamma_7} \simeq 0$, respectively.

We qualitatively study the reduction in the magnetic moment due to the hybridizations between the $d$ and $p$ orbitals on the basis of a cluster model consisting of $d$ and $p$ orbitals on oxygens around d orbitals.
In NaOsO$_3$ and SrTcO$_3$, the electronic states around the Fermi level are constructed from antibonding orbitals of $d$- and $p$-orbitals, respectively~\cite{Du,Jung,Rodriguez,Mravlje}.
The antibonding orbital $\phi_i$ is expressed as
$\phi_i =A\psi_{d_{i}} + \sum_{j=x,y,z}B_j\psi_{p_{j}}$ ($i=xy,yz,zx$), where $\psi_{d_{i}}$ and $\psi_{p_{j}}$ ($j=x,y,z$) are the wave functions of the $d_{i}$ and $p_{j}$ orbitals, respectively.
The effective SOI and effective ferromagnetic exchange interaction between $\phi$'s are $\tilde{\zeta} \simeq A^2 \zeta$ and $\tilde{J_d} \simeq A^4 J_{d}$, respectively, where we neglect the SOI on $p$ orbitals and the Coulomb interactions except for $J_{d}$.
Here, we compare the magnetic moment for $A=1.0$ (no hybridization) and that for $A=0.7$, which are reasonable values, to understand the role of hybridization between $d$- and $p$-orbitals. 
The magnetic moments are about $M_{\rm tot}/\mu_{\rm B} \simeq 2.8$ for $A =1.0$, $J_d=0.5$ (eV), and $\zeta=0.5$ (eV), and about $M_{\rm tot}/\mu_{\rm B} \simeq 1.5$ for $A =0.7$, $\tilde{J}_d \simeq A^4 J_{d} \simeq 0.1$ (eV), and $\tilde{\zeta} \simeq A^2 \zeta \simeq 0.3$ (eV).
It is found that the magnetic moment becomes smaller with the hybridization between $d$- and $p$-orbitals.  

Finally, we compare the theoretical and experimental results.
The Os ion in NaOsO$_3$ and the Tc ion in SrTcO$_3$ have (5$d)^3$ and (4$d)^{3}$ electrons, and magnetic moments of about 1$\mu_{\rm B}$~\cite{Calder} and about 1.9$\mu_{\rm B}$~\cite{Rodriguez}, respectively.
It is known that the energy splittings between $J=4$ and $J=3$ are about $E(J=4)- E(J=3) = 4\lambda_{{\rm Os}} \simeq -0.516$ (eV) for the Os atom (5$d^6$6$s^2$) and $E(J=4)- E(J=3) = 4\lambda_{{\rm Tc}} \simeq -0.0937$ (eV) for the excited state of the Tc atom (4$d^6$)~\cite{Kramida}, where $\lambda$ is the amplitude of the SOI between {\bf L} and {\bf S} determined by Hund's rule.
$\zeta$ and $\lambda$ are connected by the relation $\zeta = -(4l +2 -N)\lambda$, where $N$ and $l$ are the number of $d$ electrons and the orbital angular momentum quantum number, respectively.
The SOI $\zeta$ values for Os and Tc are estimated as $\zeta_{{\rm Os}} \simeq 0.516$ (eV) and $\zeta_{{\rm Tc}} \simeq 0.0937$ (eV), respectively, for $N=6$ and $l=2$.
In Fig. \ref{Fig4}, the magnetic moments are about 1$\mu_{\rm B}$ for $J_d=0.1$ (eV) and $\zeta=0.5$ (eV), and about 2$\mu_{\rm B}$ for $J_d=0.1$ (eV) and $\zeta=0.1$ (eV).
These results are consistent with the experimental results.
This implies that NaOsO$_3$ and SrTcO$_3$ are located in the middle of the L-S and J-J coupling schemes. 

In summary, we have discussed the effect of SOI on transition-metal ions in which three electrons occupy t$_{2{\rm g}}$ orbitals.  
We have shown that the magnetic moment of $d$ electrons on (5$d)^3$ and (4$d)^3$ is reduced by SOI.
We have qualitatively clarified that the hybridization between $p$- and $d$-orbitals assists in the reduction in the magnetic moment due to SOI. 
We have found that the (4$d)^3$ and (5$d)^3$ systems are located in the middle of the L-S and J-J coupling schemes.
 
\section*{Acknowledgment}
One of us (H.M.) would like to thank Dr. K. Yamaura, Dr. T. Kariyado, and Professor M. Ogata for valuable discussions and comments. We are grateful to Professor M. Takao for directing our attention to the present problem.

\end{document}